\documentclass[journal=nalefd,manuscript=article]{achemso}
\usepackage[version=3]{mhchem} 
\usepackage{bm}
\pdfoutput=1

\author{Nadanai Laohakunakorn}
\affiliation[University of Cambridge]
{Cavendish Laboratory, University of Cambridge, Cambridge CB3 0HE, United Kingdom}
\altaffiliation{These authors contributed equally to the work.}
\author{Benjamin Gollnick}
\affiliation[CSIC]
{\\ Centro Nacional de Biotecnolog\'{i}a, CSIC, Darwin 3, Campus de Cantoblanco, 28049 Madrid, Spain}
\altaffiliation{These authors contributed equally to the work.}
\author{Fernando Moreno-Herrero}
\affiliation[CSIC]
{\\ Centro Nacional de Biotecnolog\'{i}a, CSIC, Darwin 3, Campus de Cantoblanco, 28049 Madrid, Spain}
\author{Dirk G. A. L. Aarts}
\affiliation[University of Oxford]
{\\ Department of Chemistry, Physical and Theoretical Chemistry Laboratory, University of Oxford, South Parks Road, Oxford OX1 3QZ, United Kingdom}
\author{Roel P. A. Dullens}
\affiliation[University of Oxford]
{\\ Department of Chemistry, Physical and Theoretical Chemistry Laboratory, University of Oxford, South Parks Road, Oxford OX1 3QZ, United Kingdom}
\author{Sandip Ghosal}
\affiliation[Northwestern University]
{\\ Department of Mechanical Engineering and Engineering Sciences and Applied Mathematics, Northwestern University, Evanston Illinois 60208, United States}
\author{Ulrich F. Keyser}
\email{ufk20@cam.ac.uk}
\phone{+44(0)1223 337272}
\affiliation[University of Cambridge]
{Cavendish Laboratory, University of Cambridge, Cambridge CB3 0HE, United Kingdom}

\title{A Landau-Squire Nanojet}

\begin{document}
\newpage
\begin{abstract}
Fluid jets are found in nature at all length scales from microscopic to cosmological. Here we report on an electroosmotically driven jet from a single glass nanopore about 75 nm in radius with a maximum flow rate $\sim$ 15 pL/s. A novel anemometry technique allows us to map out the vorticity and velocity fields that show excellent agreement with the classical Landau-Squire solution of the Navier Stokes equations for a point jet. We observe a phenomenon that we call flow rectification: an asymmetry in the flow rate with respect to voltage reversal. Such a nanojet could potentially find applications in micromanipulation, nanopatterning, and as a diode in microfluidic circuits.
\newline
\newline
KEYWORDS: Nanojet, microfluidics, nanopore, flow rectification
\end{abstract}

\newpage
In 1944 an exact solution of the Navier-Stokes equations of fluid mechanics representing
a jet from a point source was found by Landau\cite{landaupapers,landaulifshitz,landau44} and later, independently by  Squire\cite{squire51}. The Landau-Squire solution, which is based on the concept of self-similarity, is one of the few exact solutions available for the nonlinear equations of fluid flow. It has found applications in a wide variety of contexts, from astrophysical jets to the splashing of liquids. 

Small scale fluidic jets have numerous applications such as inkjet printing, electrospinning, nanofabrication, drug and gene delivery, cell sorting, and microsurgery\cite{whitesides04,eggers08}. Such tiny jets can be generated by flow focusing using thermal, piezoelectric, or electrohydrodynamic techniques \cite{basaran02}. Focusing using a concentric coflowing stream \cite{ganancalvo07,knight98} can create jets with diameters of a few hundred nanometers. We create a nanojet by simply driving fluid through a nanopore using an electric field.

Nanopores have been widely used as single-molecule sensors to probe the structure and conformation of biomolecules such as DNA. Solid-state nanopores based on silicon nitride \cite{dekker07} and glass \cite{steinbock10}  generally have a negative surface charge which is screened by a Debye layer of positive counterions when the pore is filled with an electrolyte. When an electric field is applied along the pore axis, the screening layer of counterions moves in response to the applied field, driving the rest of the fluid in the channel through viscous coupling. This electroosmotic flow is known to have a strong influence on the translocation properties of biomolecules through nanopores \cite{vandorp09, ghosal07, ghosal07_prl, laohakunakorn13}. In addition, by tuning the inner surface of the pore, the flow rate and direction can be controlled \cite{sparreboom09, bouzigues08}.

Micro- and nanofluidic flow measurements are challenging because of the small flow rates involved. Techniques that have been developed to overcome these problems include measurements using electrical admittance \cite{collins04}, current monitoring \cite{huang88}, thermal transport \cite{ernst02}, droplet size monitoring \cite{sinha07}, and periodic flapping of a jet \cite{lee02}. Particle image velocimetry (PIV) \cite{santiago98} adapted to small scale systems can yield detailed information on the spatial distribution of velocities. More recent techniques include laser induced fluorescence photobleaching anemometry \cite{wang05, kuang10} and electric cross-correlation spectroscopy \cite{mathwig12}. One of the challenges in nanoscale measurements of electroosmotic flow is that tracer particles are almost invariably charged and respond to the electrophoretic force in addition to the hydrodynamic drag. The particle's observed velocity may therefore not equal the local flow velocity even if particle inertia is negligible.  Thus, PIV measurements for electroosmotic flows have to be carried out using neutral tracer particles, which poses an additional set of challenges due to particle aggregation.

We have developed a novel method for characterizing the flow field around electroosmotically driven jets from nanopores emerging into a buffer of stationary fluid; we exploit the well known result from fluid mechanics that a small particle embedded in a flow rotates at an angular velocity equal to half the local vorticity~\cite{brenner83}. Our nanopores are fabricated by heating glass capillaries with a focused laser and pulling using a commercial pipet puller until they break (see Materials and Methods). Our setup is shown in Figure~1A,B. We use a coordinate system where the origin is at the pore, the $x$-axis is directed along the axis of the jet, and the $z$-axis is parallel to
the optical axis of the microscope (Supporting Information Figure~S1). The rotation of the particle can be detected as periodic modulations of the scattered light if it has a slight asymmetry that displaces the center of its optical image from the center of rotation. We use custom-made colloidal microparticles shaped like dimpled spheres, which exhibit the required asymmetry (Figure~1C; see Materials and Methods). The particle is held in a laser optical trap and positioned in front of a 75 nm-radius glass nanopore as depicted in Figure~1A,B using a piezoelectric nanopositioning system (see Materials and Methods). The pore and reservoir are filled with an aqueous salt solution (10 mM KCl). A voltage is applied between the inlet and outlet reservoirs resulting in an electroosmotic nanojet through the pore.

The position of the optical center of the particle is tracked with 2-nm accuracy at a few kilohertz using a high-speed CMOS camera \cite{otto10}. When the particle is subject to a shear flow it rotates (see Supporting Information Movie~S2). Since the optical center is displaced relative to the center of rotation, we observe periodic fluctuations in the apparent position $(X,Y)$ of the particle relative to the trap center. The phase difference between the two signals is close
to $\pi/2$ (Figure~1D) and indeed, $(X,Y)$ traces out a roughly circular trajectory, as shown in Figure~1E.

In order to characterize the nanojet we move the particle in the $x$-$y$ plane. For each $x$- location, the particle rotation frequency is measured at a number of $y$- locations that allows us to construct a map representing the flow field. One advantage of this approach is that the particle rotation is sensitive only to the hydrodynamic field. This is because the torque experienced by the particle is not sensitive to its charge. Although the slight asymmetry of the particle is essential for determining its rotation rate from light scattering, this does not significantly alter its mechanical properties; consequently, we treat it as a sphere (radius $a$) embedded in a flow (for a more complete exposition of the theory please see the discussion in Materials and Methods). The force and torque on the particle are then given by
Faxen's laws \cite{brenner83} from which it follows that the rotation rate ($\bm{\Omega}$) is related to the fluid vorticity ($\bm{\omega}$) as
$\tau  \dot{\bm{\Omega}} =   \bm{\omega}/2  - \bm{\Omega}$. Here $\tau =(\rho_{c} a^{2}/15 \mu)$ is an equilibration time scale,
$\rho_c$ is the density of material of the particle, and $\mu$ is the viscosity of water. In our situation,
$\tau \sim 0.1$ $\mu$s, so that the particle can be taken as rotating at its equilibrium ``terminal" rotation frequency $\bm{\Omega} = \bm{\omega}/2$.

The Landau-Squire (LS) solution corresponds to the limiting situation where the radius ($R$) of the jet emerging into the quiescent fluid approaches zero but the fluid momentum flux ($P$) is held constant~\cite{landaupapers,landaulifshitz,landau44,squire51}. If the cross-sectional variation of the flow velocity at the pore is neglected, then $P = Q_{0} \rho_{0}  u_{0} = \rho_{0} Q_{0}^{2}/(\pi R^{2})$, where the volumetric flux $Q_0=\pi R^2 u_0$, $u_0$ is the mean velocity at the pore exit, and $\rho_0$, the fluid density. The jet Reynolds number in our experiment is $\text{Re} = [ \rho_{0} P/(2 \pi \mu^{2}) ]^{1/2} \sim 0.05$ so that we are only concerned with the low Reynolds number limit of the LS jet. In this limit, the flow is identical to a ``Stokeslet", the flow generated in a quiescent Stokesian fluid under the action of a point  force. The stream function for such a flow is $\psi (r, \theta) = (P/8 \pi \mu) r \sin^{2} \theta$ where $(r,\theta,\phi)$ are spherical coordinates centered on the pore with $\theta$ measured from the direction of the axis pointing into the outlet reservoir (Supporting Information Figure~S1). Thus, the particle rotation frequency is given by $\bm{\Omega} = \bm{\omega}/2 = P\sin \theta \hat{\bm{\phi}} /(8\pi \mu r^2)$, where $\hat{\bm{\phi}}$ is the unit vector in the azimuthal direction. 

The method of measurement of the particle rotation frequency is summarized in Figures~1 and 2A. Data from one experiment displaying the variation of rotation frequency  ($\Omega$) with transverse location ($y$) of the particle for a fixed axial distance ($x$) from the pore is shown in Figure~2B. Since $\Omega = P\sin \theta  /(8\pi \mu r^2) = (P/8 \pi \mu) y^{*}$ where $y^{*} \equiv y / (x^{2} + y^{2})^{3/2}$ we expect $\Omega$ to increase with $y$ for small $y$ and decrease with $y$ for large $y$ which is consistent with the qualitative shape of the rotation curves seen in Figure 2B. In order to make a quantitative comparison we replot the same data in Figure~3A by plotting $\Omega$ as a function of the variable $y^{*}$, where for each voltage, measurements were made at three different axial locations. We see that $\Omega$ depends on $x$ and $y$ only through the combination $y^{*}$ and $\Omega \propto y^{*}$, as required by the LS solution. The slope of this straight line then gives us $P = \rho_{0} Q_{0}^{2}/(\pi R^{2})$. Since the pore radius is known from SEM measurements (see Materials and Methods), the flux $Q_{0}$ may be obtained.

In addition to the rotation frequency obtained from the oscillating signal, we are able to measure the force vector ($\bm{F}$) based on the linear displacement of the particle in the optical trap. This is related to the fluid velocity ($\bm{v}$) at the particle location in the absence of the particle as $\bm{v} = \bm{F}/(6 \pi \mu a)$ (see discussion in Materials and Methods). The measured force vectors at selected spatial locations are shown in Figure~2C. From the LS solution, this force is $F=3Pa/(2 r^{*})$ where $r^*=r / [ \cos^2\theta+\sin^2\theta/4 ]^{1/2}$. The rescaled data is shown in Figure~3B which confirms the proportionality between  $F$ and $1/r^*$. Measuring the slope provides a way of determining the momentum flux $P$ and hence the flow rate $Q_{0}$ independent of the rotation measurements. This second method has the disadvantage that unlike the rotation frequency, the force could be sensitive to electric fields. In the present experiment, however, this is not an issue as the electric field drops off very rapidly from the pore exit. At distances of the order of several pore diameters where we locate our particle the electric field is essentially reduced to zero (Supporting Information Figure~S3). Indeed, the force on the particle is not substantially changed if the particle is coated to increase, decrease or reverse its charge (Supporting Information Figure~S4).

It should be noted that the observed rotation is due to hydrodynamic shear and not due to the mechanism of ``Quincke rotation" often discussed in the context of the electromechanics of small particles~\cite{quincke96, jones84, das13, jones95}. Indeed, the critical field for the onset of Quincke rotation in our case is about $10^{8}$ V/m -- almost 6 orders of magnitude higher than the typical electric field at the location of the colloid (Supporting Information Figure~S3). Furthermore, since the angular velocity of Quincke rotation is determined by the magnitude of the electric field, the speed of rotation of the particle should be maximum on the centerline and decrease monotonically with transverse displacement. Instead, the observed rotation speed (Figure~2B) is zero on the axis and first increases and then decreases on transverse displacement away from the axis. These observations are consistent with shear induced rotation due to an electroosmotic jet from the pore but not consistent with rotation driven by the Quincke mechanism.

The flow rates determined from particle rotation as well as from the force measurements are shown in Figure 4 for a range of applied voltages. We see that the results using both methods of measurement agree to within experimental errors. The flow rates observed are of the order of a few tens of pL/s, which are at the sensitivity limit of most other flow measurement techniques \cite{kuang10, collins04, ernst02, mathwig12}. The results shown here were obtained using a 1.5 $\mu$m-diameter particle; however the measured flow rates are independent of particle size (Supporting Information Figure~S5). 

The dependence of the current ($I$) through the pore on the applied voltage ($V$)  is also shown in Figure~4. We see that the function $I(V)$ is not antisymmetric, a property that is known as ``current rectification" \cite{siwy06}. We find a similar behavior for the flow rate $Q(V)$, except it works in opposition to current rectification: when the ionic current is high the fluid flow is low and vice versa (Supporting Information Table S6). We call this striking effect ``flow rectification". A similar feature was reported in a pyramidal-pore membrane \cite{jin10}, but here we show this for a single nanopore. Because of the linearity of the Navier-Stokes equations at low Reynolds numbers, reversing the applied voltage reverses the flow field. This feature of low Reynolds number flows has traditionally made it difficult to design microfluidic flow rectifiers; such devices have either taken advantage of geometries that increase the Reynolds number so that inertial effects become important \cite{nabavi09, tesla20} or introduced nonlinearities into the system, for instance by making the liquid non-Newtonian \cite{sousa10, groisman04}. In our system, the nonlinearity may be due to a local departure from the Stokes flow limit very close to the pore entrance where the fluid undergoes significant acceleration or perhaps due to nonlinear electrokinetic phenomena such as induced-charge electroosmosis. The exact mechanism for the rectifying behavior is however unclear at the present time.

The electrokinetic nanojet described here has a number of potential novel applications. Possibilities include use as a ``flow rectifier" in microfluidic logic circuits, the functional equivalent of semiconductor diodes in microelectronics, and also in applications involving nano scale patterning and micro manipulation\cite{bruckbauer02}. 

Finally, we note that notwithstanding the nanometer size of the jet and the picoliter-per-second flow rates, 
the classical continuum theory of Landau and Squire appears to work remarkably well. It is interesting to 
draw a comparison with a recent molecular dynamics study \cite{moseler_landman_13} of a nanometer sized propane jet emerging into a vacuum. Here too, good agreement with continuum theory was found except for the fact that fluctuations could play a dominant role in certain processes such as jet break up. Our work differs from this earlier study in that we demonstrate here a physical nanojet rather than a numerical one, and, our flow is due to a fluid jet emerging into a reservoir of the quiescent fluid and not a free surface flow showing instabilities and break up.

\section{Materials and Methods}
\subsection{Experimental Procedures}
Our combined optical tweezers and nanopore setup is based on a custom-built inverted microscope, and has been described previously \cite{otto10, steinbock10_jphys}. Briefly, a 1,064-nm ytterbium fiber laser is focused using a high-NA objective to a diffraction limited spot which traps micrometer-sized dielectric particles stably in three dimensions. Our nanopores are fabricated by pulling quartz glass capillaries using a programmable commercial laser puller (P-2000, Sutter Instruments).
The nanopores generated in this way can have radii as small as tens of nanometers \cite{steinbock10, steinbock10_jphys}.
The data reported here were obtained with a batch of capillaries with pore radii in the range 74$\pm$13 nm, as measured by imaging in an SEM. The pore is assembled into a microfluidic chip where it connects two reservoirs filled with a salt solution (10 mM KCl, 1 mM Tris-EDTA, pH 8). Ag/AgCl electrodes connect the reservoirs to a commercial patch-clamp amplifier (Axopatch 200B, Axon Instruments), which allows for application of voltages between -1 and +1 V as well as low-noise ionic current monitoring through the pore. Asymmetric dimpled colloidal particles (diameters 1.5 $\mu$m, 2 $\mu$m, and 3 $\mu$m) made of 3-methacryloxypropyl-trimethoxysilane (TPM), with a mass density of 1.228 g/mL, were synthesized according to the method described previously~\cite{sacanna10} . These are flushed into the cell before a single particle is trapped. Position detection within the trap is achieved using a high-speed CMOS camera (MC1362, Mikrotron) running at several kilohertz and processed using custom LabVIEW software. The trap stiffness is calibrated using a power spectral density method. The position of the trap with respect to the pore is controlled using a piezoelectric nanopositioning system (P-517.3 and E-710.3, Physik Instrumente). The application of voltage generates a flow that causes the colloid to rotate, and this rotation is measured directly using the position detection software.

\subsection{Theory}
\subsubsection*{1. Relation between Vorticity and Particle Rotation}
An infinitesimal fluid element in a flow rotates at an angular velocity $\bm{\Omega}=\bm{\omega}/2$ where $\bm{\omega}$ is the vorticity vector. If a small spherical solid particle (radius $a$) is embedded in a flow, the force and torque on the particle are given ~\cite{brenner83} by Faxen's laws:
\begin{equation}
\bm{F}=6\pi\mu a(\bm{v}-\bm{U})+\pi\mu a^3\nabla^2\bm{v}
\label{eqn:fax1}
\end{equation}
\begin{equation}
\bm{T}=8\pi\mu a^3\left(\frac{\bm{\omega}}{2}-\bm{\Omega}\right).
\label{eqn:fax2}
\end{equation}
Here, $\bm{v}$ is the fluid velocity of the ambient flow in the absence of the particle, $\bm{\omega}=\bm{\nabla}\times \bm{v}$ is the vorticity, $\bm{U}$ and 
$\bm{\Omega}$ are respectively the linear and angular velocities of the particle, and $\mu$ is the dynamic viscosity of the fluid. By ``small particle'" we mean that $a\ll L$, $L$ being a characteristic length over which the flow varies. In our experiment this requirement of ``smallness" is only marginally satisfied. Nevertheless,
measurements do not show a marked dependence of our results on colloid size (see Supporting Information Figure~S5) indicating that any correction 
due to finite particle size is probably small. 

Consider the rotation of a colloidal particle tethered in the flow by the optical trap (which does not exert a torque on the particle). The equation for the angular velocity of the particle is then
\begin{equation}
I \frac{d\bm{\Omega}}{dt}=8\pi\mu a^3\left(\frac{\bm{\omega}}{2}-\bm{\Omega}\right)
\label{eqn:motion}
\end{equation}
where $I=(2/5)ma^2$ is the moment of inertia of a sphere of mass $m$. Even though the fact that the particle is distorted from a spherical shape is important for the detection of the rotation rate, we assume that this departure from sphericity has a negligible effect on the particle dynamics. If the particle's material density is $\rho_c$, eq 3 may be written as
\begin{equation}
\tau\frac{d\bm{\Omega}}{dt}=\frac{\bm{\omega}}{2}-\bm{\Omega}
\end{equation}
where
\begin{equation}
\tau=\frac{\rho_{c} a^2}{15\mu}
\end{equation}
is an equilibration time scale. It is the characteristic time in which a spinning particle will be brought to rest due to fluid viscosity if the fluid is still, or equivalently, the time scale on which the particle rotation rate equilibrates with the local shear.

\subsubsection*{2. The Landau-Squire Jet}
Consider a jet emerging from a pore (radius $R$) with an average velocity $\bm{u}_{0}$ . The mechanism driving the jet (i.e. whether electrokinetic or pressure driven) is of no consequence for the discussion that follows. This jet has a volume flux $Q_0=\pi R^2u_0$ and a momentum flux $P=\pi R^2 \rho_0 u_0^2$. If we consider the limit $R\rightarrow0$ but $P$ fixed, such a limiting situation is referred to as a \emph{jet from a point source of momentum}. Notice that this limit implies that $Q_0=\pi R^2u_0=R\sqrt{\pi P/\rho_0}\rightarrow0$. Thus, the jet may be regarded as a flow due to a point force $P$ within the body of the fluid that accelerates the fluid locally. The volume flux across a cross-section at a distance $x$ from the orifice, $Q(x)$, is however not zero but in fact increases with $x$. This is because, the jet entrains fluid from the surrounding stagnant pool so that the fluid carried in the jet increases indefinitely. For such an idealized jet, $Q(0)=Q_0=0$, thus, the jet adds 
momentum but no mass to the surrounding fluid. 
The jet Reynolds number ($\text{Re}$) is defined as
\begin{equation}
\text{Re}=\left(\frac{\rho_0 P}{2\pi\mu^2}\right)^{1/2}.
\end{equation}
Inserting characteristic values, we have $\text{Re}\sim0.05$.

The jet from a point source of momentum is one of a handful of cases where an exact solution of the nonlinear Navier-Stokes equation is available.
In the limit of low Reynolds number, the stream function of the flow in spherical polar coordinates centered at the pore (see Supporting Information Figure~S1) is
\begin{equation}
\psi(r,\theta)=\frac{P}{8\pi\mu}r\sin^2\theta.
\end{equation}
The velocity components can be obtained from the stream function as follows:
\begin{align}
u_r&=\frac{1}{r^2\sin\theta}\frac{\partial\psi}{\partial\theta} = \frac{P\cos\theta}{4\pi\mu r} \label{eqn:ur}\\
u_\theta&=-\frac{1}{r\sin\theta}\frac{\partial\psi}{\partial r} = -\frac{P\sin\theta}{8\pi\mu r}. \label{eqn:ut}
\end{align}
Taking the curl of this vector field yields the vorticity
\begin{equation}
\bm{\omega}=\frac{P\sin\theta}{4\pi\mu r^2}\bm{\hat{\phi}},
\end{equation}
where $\bm{\hat{\phi}}$ is the unit vector in the azimuthal direction. 

\subsubsection*{3. Extracting the Flow Rate from Measurements}
On expressing the Landau-Squire solution in Cartesian coordinates (as defined in Supporting Information Figure~S1), we find that the angular velocity of the colloid is 
$\bm{\Omega}=\bm{\omega}/2=\Omega\bm{\hat{\phi}}$, where
\begin{align}
\Omega&=\frac{P}{8\pi\mu}\frac{\sin\theta}{r^2}=\frac{P}{8\pi\mu}y^* \label{eqn:omega} \\
y^*&\equiv\frac{y}{(x^2+y^2)^{3/2}}.
\end{align}
Equation 11 predicts that $\Omega$ should increase linearly with $y$ for small  (relative to the distance from the pore, $x$) $y$, and decrease as $|y|^{-2}$ for large $|y|$ . It follows that the curve $\Omega(y)$ would have a maximum (or minimum) for some intermediate $|y|\sim x$. This behavior is in qualitative agreement with the rotation curves shown in Figure~2B. A more quantitative comparison may be made by plotting $\Omega$ as a function of $y^*$ which should yield a straight line. The 
momentum flux $P$, and hence flow rate $Q_0$ can then be extracted from the slope of the line.

From eqs 8 and 9, the magnitude of the velocity vector at a location $(r,\theta)$ is given by
\begin{align}
u&=\sqrt{u_r^2+u_{\theta}^2} \\
&=\frac{P}{4\pi\mu r}\sqrt{\cos^2\theta+\frac{\sin^2\theta}{4}}.
\end{align}
Thus, for a small particle ($a \ll L$) that is tethered ($\bm{U}=0$), eq 1 implies that 
\begin{align}
F&=6\pi\mu a u \\
&= \frac{3Pa}{2r}\sqrt{\cos^2\theta+\frac{\sin^2\theta}{4}} \\
&=\frac{3Pa}{2r^*}
\end{align}
where 
\begin{equation}
\frac{1}{r^*}=\frac{1}{r}\sqrt{\cos^2\theta+\frac{\sin^2\theta}{4}}.
\end{equation}
Thus, a plot of $F$ against $1/r^*$ should yield a straight line from the slope of which the flow rate may be determined.
This provides us with a second method of measuring the flow rate independent of the colloid rotation measurements.

\begin{acknowledgement}

The authors thank Nicholas Bell for help with SEM imaging, and Christian Holm for helpful comments and discussions. N.L. is funded by the George and Lillian Schiff Foundation and Trinity College, Cambridge. B.G. is supported by an FPU Ph.D. scholarship from the Spanish Ministry of Education. S.G. acknowledges support from the NIH through grant 4R01HG004842-03 and from the Leverhulme Trust in the form of a Visiting Professorship.  U.F.K. acknowledges funding from an Emmy Noether grant (Deutsche Froschungsgemeinschaft) and an ERC starting grant. 

\end{acknowledgement}

\begin{suppinfo}
Four figures clarifying the coordinate system, showing the effects of changing colloid charge and size, and showing the calculated electric field in the nanopore; one table quantifying the flow and current rectification behavior observed; and one movie illustrating the experimental procedure.
\end{suppinfo}

\newpage 

\bibliography{bibliography}

\begin{figure}
\begin{center}
\includegraphics[]{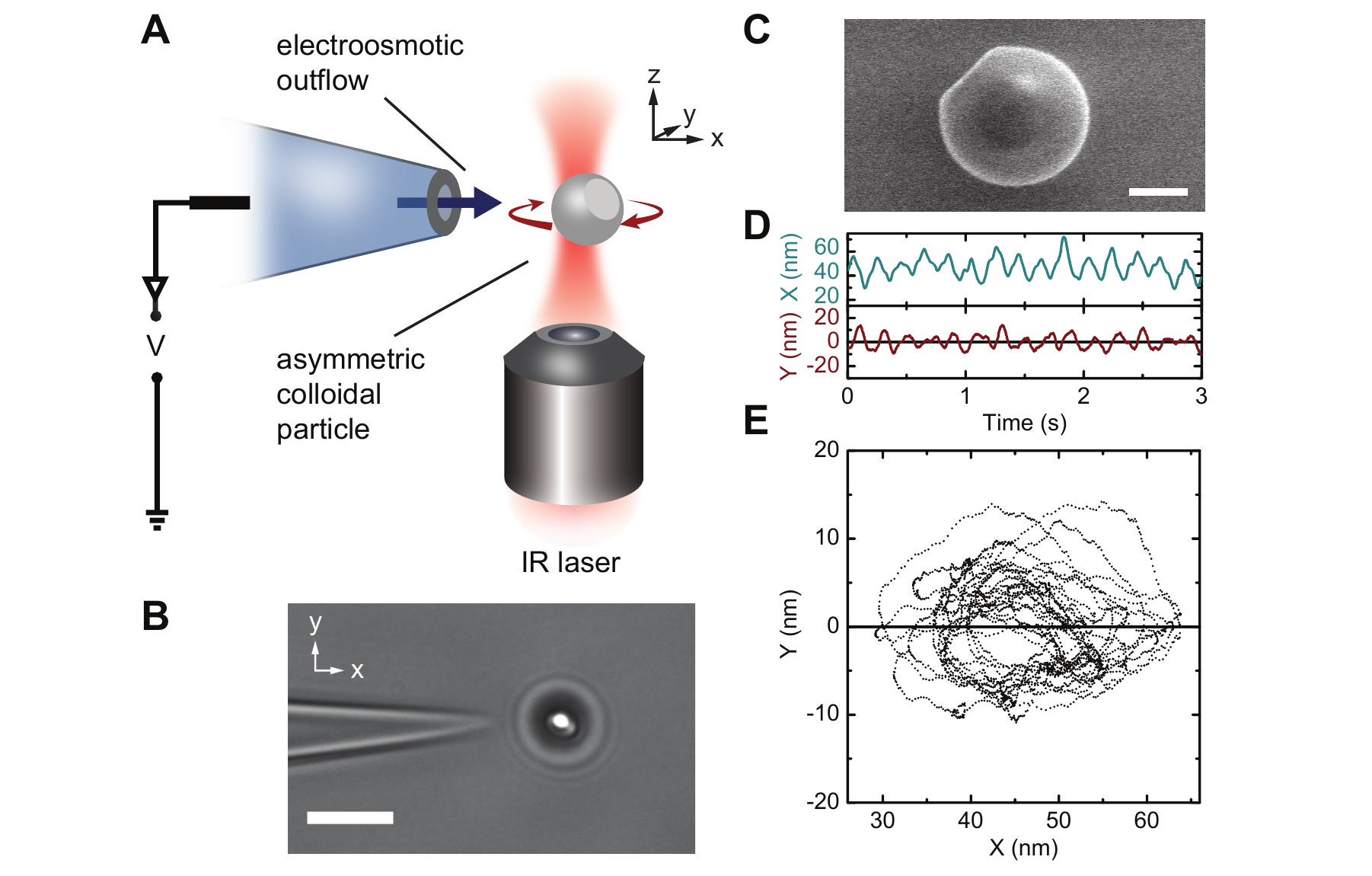}
\end{center}
\caption{\label{fig:figure1}Particle rotation anemometer based on optical tweezers. (A) A schematic of the setup (not to scale): an asymmetric dimpled colloidal particle is held in the optical trap near a conical glass nanopore of radius 75 nm. A voltage is applied across the pore and the rotation rate of the particle measured. (B) A light micrograph showing an experiment in progress. The scale bar corresponds to 5 $\mu$m. (C) An SEM micrograph of a 3 $\mu$m diameter asymmetric particle showing a clear dimple on one side. The scale bar corresponds to 1 $\mu$m. (D) The raw data of the position $(X,Y)$ of the particle relative to the trap center obtained at 1 kHz, smoothed with a window of 50 points for clarity. The $x$ direction is along the pore axis, while $y$ is transverse. The displacement in $X$ is due to a force on the particle from the fluid flow. (E) An $X$-$Y$ phase plot reveals that the optical center of the particle undergoes a roughly circular motion.}
\end{figure}

\begin{figure}
\begin{center}
\includegraphics[]{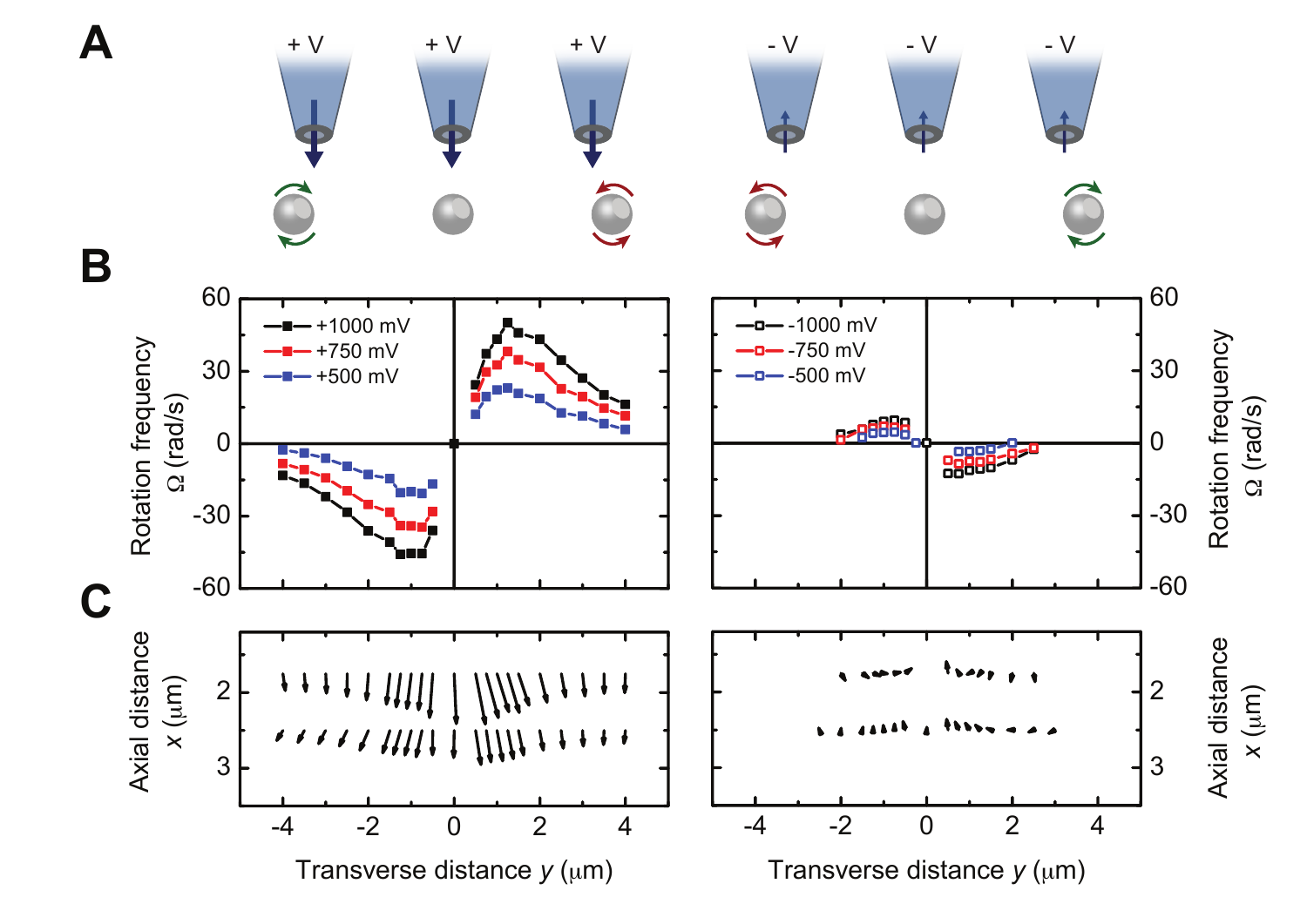}
\end{center}
\caption{\label{fig:figure2}Simultaneous measurement of rotation and force probes the flow field. (A) A schematic of the experimental procedure. (B) Scanning a 3 $\mu$m particle transverse to the pore axis (in the $y$ direction) leads to antisymmetric rotation curves as shown; as the particle is moved from far afield along the $y$-axis, the rotation rate first increases, then decreases to  zero on the jet axis, and finally, reverses direction. (C) Also shown is the vector force on the particle, for $V=+1000$ mV and $V=-1000$ mV at two different axial ($x$) distances. We observe strong outflows at positive voltages but weak inflows at negative voltages.}
\end{figure}

\begin{figure}
\begin{center}
\includegraphics[]{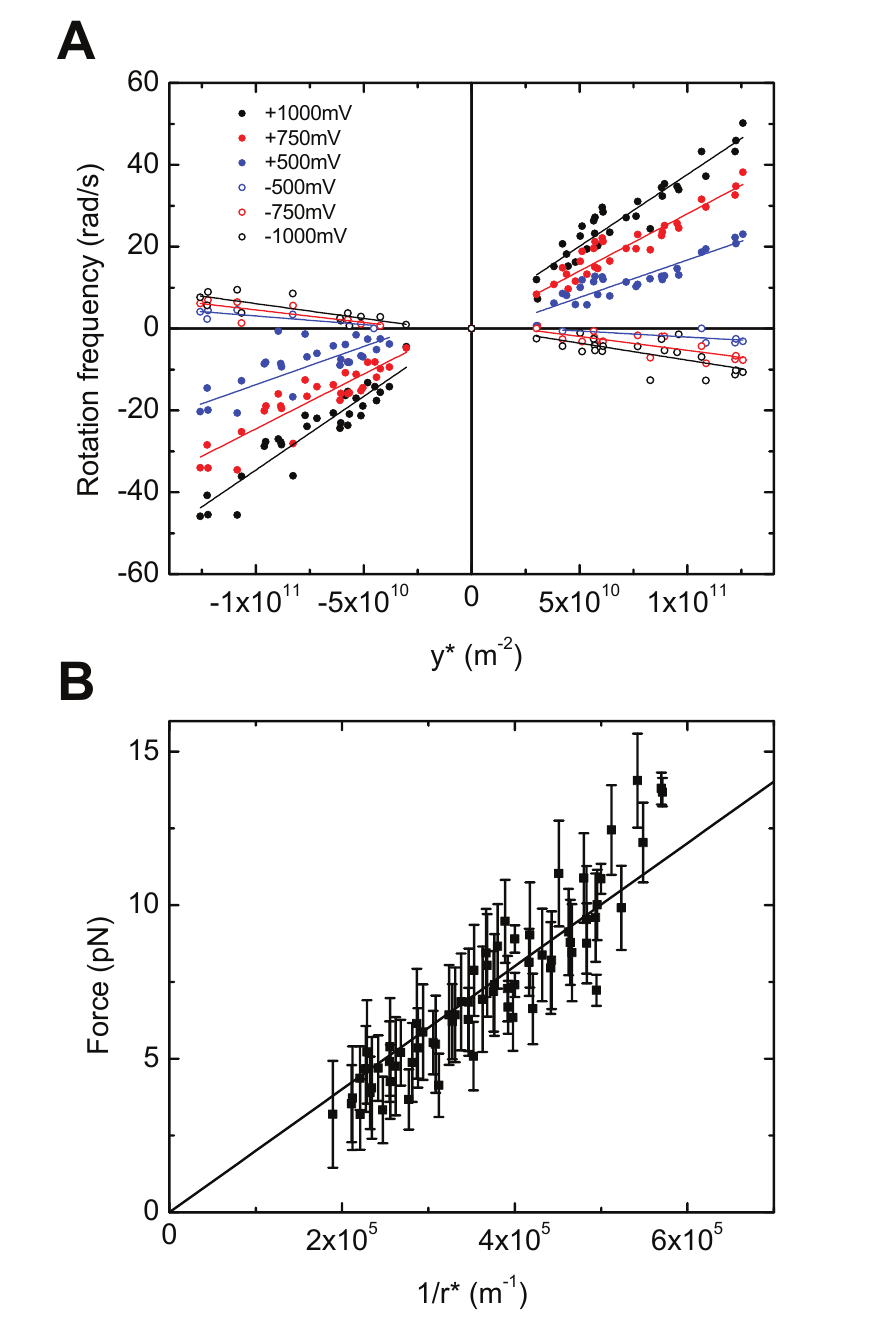}
\end{center}
\caption{\label{fig:figure3} Verification of the Landau-Squire scaling. (A) Rotation measurements were carried out at transverse positions between -4 $\mu$m $<y<$ 4 $\mu$m at each of three different axial positions, $x=1.75$, 2, and 2.5 $\mu$m. All of this data collapses on to a single straight line when plotted against the similarity variable $y^*\equiv y/(x^2+y^2)^{3/2}$. Different voltages give lines with different slopes. The flow rate can be extracted from the slope of the line, as described in the text. The data shown here was obtained using a 3 $\mu$m particle. (B) The magnitude of the force on the particle ($F$) also yields a straight line when plotted against
$1/r^*= r^{-1} ( \cos^2\theta+\sin^2\theta/4 )^{1/2}$. Here only the data set at $V=+1000$ mV is shown for clarity.}
\end{figure}

\begin{figure}
\begin{center}
\includegraphics[]{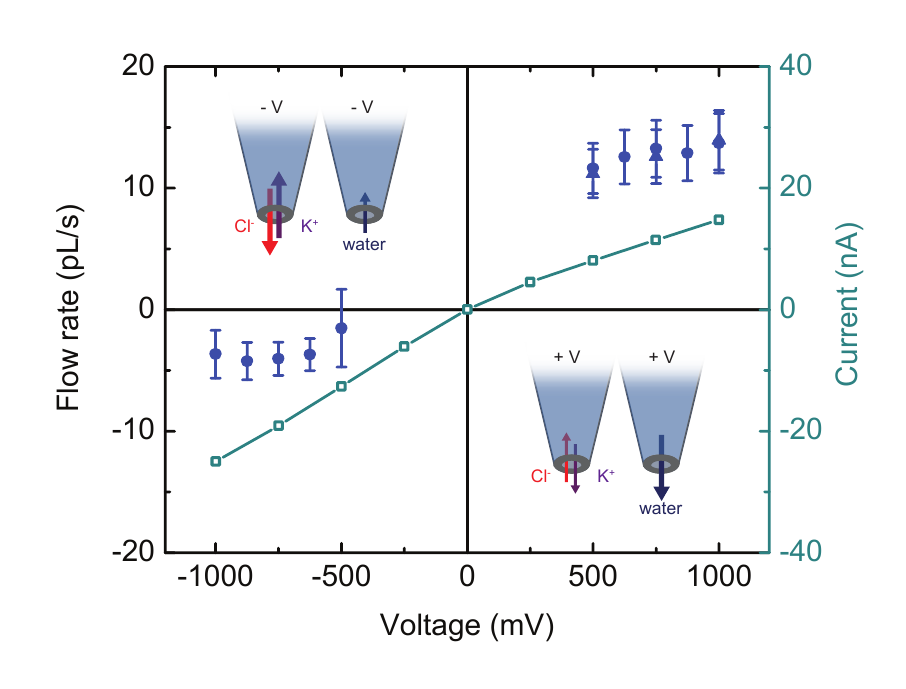}
\end{center}
\caption{\label{fig:figure4} Observation of electroosmotic flow rectification. We observe $\sim$pL/s flow rates, which exhibit a rectification behavior as the voltage is varied from -1 to +1 V. Measurements of rotation (circles) and force (triangles) give the same result within errors, which indicate the standard deviation of the measurements. Flow rectification behaves in the opposite sense to ion current rectification which is apparent from the current-voltage characteristic shown in cyan. The measurements shown in this figure were obtained using a 1.5 $\mu$m-diameter particle.}
\end{figure}

\end{document}